\documentclass[twocolumn]{aastex63}

\newcommand\etal{\it{et al.} \rm }
\submitjournal{ApJ}

\shorttitle{Spin parity III}
\shortauthors{Iye \etal}
\begin{document}

\title{Spin Parity of Spiral Galaxies III -- Dipole Analysis of  the Distribution of SDSS Spirals with 3D Random Walk Simulations }

\correspondingauthor{Masanori Iye}
\email{m.iye@nao.ac.jp}

\author{Masanori Iye}
\affil{National Astronomical Observatory of Japan, Osawa 2-21-1, Mitaka, Tokyo 181-8588 Japan}
%\affil{National Institutes of Natural Sciences, Hulic Kamiyacho Building 4-3-13 Toranomon, Minato,  Tokyo 105-0001 Japan}

\author{Masafumi Yagi}
\affil{National Astronomical Observatory of Japan, Osawa 2-21-1, Mitaka, Tokyo 181-8588 Japan}

\author{Hideya Fukumoto}
\affil{The Open University of Japan, 2-11 Wakaba, Mihama-ku, Chiba 261-8586 Japan}



\begin{abstract}

Observation has not yet determined whether the distribution of spin vectors of galaxies is truly random. It is unclear whether is there any large-scale symmetry-breaking in the distribution of the vorticity field in the universe. Here, we present a formulation to evaluate the dipole component $D_{max}$ of the observed spin distribution, whose statistical significance $\sigma_{D}$ can be calibrated by the expected amplitude for 3D random walk (random flight) simulations.  

We apply this formulation to evaluate the dipole component in the distribution of Sloan Digital Sky Survey (SDSS) spirals.  \citet{Shamir2017a} published a catalog of spiral galaxies from the SDSS DR8, classifying them with his pattern recognition tool into clockwise and counterclockwise (Z-spiral and S-spirals, respectively). He found significant photometric asymmetry in their distribution.  We have confirmed that this sample provides dipole asymmetry up to a level of $\sigma_{D}=4.00$. 

However, we also found that the catalog contains a significant number of multiple entries of the same galaxies. After removing the duplicated entries, the number of samples shrunk considerably to 45\%. The actual dipole asymmetry observed for the 'cleaned' catalog is quite modest, $\sigma_{D}=0.29$.  We conclude that SDSS data alone does not support the presence of a large-scale symmetry-breaking in the spin vector distribution of galaxies in the local universe. The data are compatible with a random distribution.

\end{abstract}

\keywords{anisotropy---catalogs---galaxies: formation---galaxies: rotation---galaxies: spiral}

%% cross-references to sections, equations, tables, \& figures.

%% \& \citet commands to identify citations.  The citations are
%% tied to the reference list via symbolic KEYs. The KEY correspondsh

\section{Introduction} 
%\label{sec:intro}
For many decades, investigation of the formation and evolution of galaxies has been primary subject of astrophysics. Semi-analytic simulations of structure formation in the $\Lambda$\rm CDM model of the universe to reproduce clustering and merging of galaxies provide the current standard picture of galaxy formation \citep{White1978, Steinmetz2002}.  \citet{Ferreira2020} suggests that ultralight dark matter may reconcile certain remaining problems which the standard models fail to explain. Recent high-spatial resolution simulations enable tracking of the formation of the spiral structure of individual galaxies \citep{Robertson2004, Agertz2011, Ceverino2017, Shimizu2019}.

On the other hand, there are other classical models of galaxy formation, such as the primordial whirl scenario \citep{Weizsaecker1955}, the pancake shock scenario \citep{Peebles1969, Zeldovich1970}, and the tidal torque scenario \citep{White1984}.  Each provides naive inference regarding the statistical distribution of the spin vectors of galaxies. 

If the spin vectors of individual galaxies were produced by splitting of large-scale primordial whirls, there would be some remaining coherent spin alignment parallel to the primordial whirl vectors, producing an observable dipole anisotropy. 

If the galaxies were formed mainly at the equatorial planes of primordial collapsing pancakes, spin vectors can mostly be expected to be parallel to the equatorial planes, possibly producing a quadrupole anisotropy seen from the observer located within a cluster of galaxies.

If an individual galaxy started spinning owing to the tidal torque from a galactic cluster mass assembly, the galaxy's spin vector should initially be perpendicular to the line joining the galaxy and the center of gravity of the cluster mass assembly. Such an initial correlation would, however, be diluted soon by orbital mixing.

\begin{figure}[ht]
\centering
\includegraphics[scale=0.85]{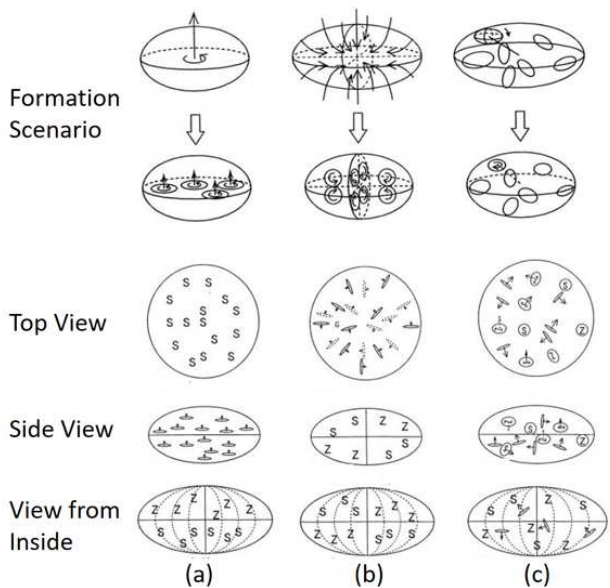}
\caption{Three illustrative scenarios of galaxy formation: (a) primordial whirl, (b) pancake shock, and (c) tidal torque. Inferred distributions of spin vectors of galaxies are shown in top views, side views, and views for observers located inside the cluster of galaxies (modified from a figure in \citet{Sugai1995}).}
\label{fig1}
\end{figure}

Figure~\ref{fig1} illustrates potential anisotropy that might be induced from different galactic formation scenarios \citep{Sugai1995}.
While the orbital mixing and merging of galaxies could wipe out these initial spin vector anisotropy, if any, observational verification of any symmetry-breaking in the distribution of spin vectors would be a significant evidence. 

There were hot debates in the middle of the last century over whether the spiral structure of galaxies winds in a trailing way or in a leading way. It has been generally believed that the spiral structure of galaxies is in general "trailing" rather than "leading."  \citet{Iye2019} found corroborative evidence from their survey of 146 spiral galaxies that all spiral structure in these galaxies is indeed "trailing."  This confirmation provides a basis for us to use the spiral winding direction projected on the sky, either S-wise spirals or Z-wise spirals, to judge the sign of the line-of-sight component $\Omega_{d}$ of a galaxy's spin vector. This determines whether the spin vector is pointing away or toward us.

\citet{Borchkhadze1976} pointed out the number dominance of "S"spirals over "reverse S" spirals, hereafter referred as Z-spirals in the present paper, in the 7,563 galaxies sampled from the Abastumani catalog of Bright Galaxies.
\citet{Mac1985} found a similar trend and a marginal dependence on the super galactic hemisphere. 
\citet{Iye1991} and \citet{Sugai1995} used S/Z parity information on spiral galaxies to study the distribution of spin angular momentum vectors in the assembly of galaxies. However the data sets available at that time were not large enough. 

The Galaxy Zoo 1 catalog, a morphological classification of SDSS galaxies, was compiled by a public poll. It showed a similar predominance of S spirals as previous studies, but \citet{Hayes2017} interpreted this as being due to human selection bias rather than a human chirality bias or physical reality.   

\citet{Shamir2017a} published tables of 82,244 clockwise (Z-spirals) and 80,272 counterclockwise (S-spirals) galaxies by using his \it{Ganalyzer} \rm algorithm \citep{Shamir2011a, Shamir2011b} for a dataset of 740,908 galaxies classified as spiral galaxies from three million SDSS Data Release 8 galaxies \citep{Kuminski2016}. The statistics show over-dominance of Z-spirals from a random distribution at $3.46\sigma$. \citet{Shamir2017b} reported finding a significant "photometric anisotropy" such that the mean magnitudes of S/Z spirals behave differently depending on their right ascension with a possible asymmetry axis at ($\alpha = 172^{\circ}, \delta = +50^{\circ}$) in J2000. This axis corresponds to the direction ($l = 152^{\circ}, b = +62^{\circ}$) near the galactic pole ($b=+90^{\circ}$). Note that Shamir's catalog shows dominance of Z-spirals instead of the S-spiral dominance reported by previous workers.

In the present paper, we develop a formulation to analyze the dipole anisotropy of S/Z spin distribution of galaxies and apply this formulation to reanalyze the Shamir's SDSS catalog.

\section{Formulation of Dipole Analysis}

\subsection{Dipole Component of Spin Distribution}

Number statistics can be used to study the S/Z number asymmetry from equipartition. The significance level of the number asymmetry is given by 
\begin{equation}
\sigma_{N}=|N(S)-N(Z)|/\sqrt{N(S)+N(Z)}.  
\end{equation}

Another option is to look for any statistically significant dipole in the spatial distribution of S/Z-spirals.

Let $\mathbf{\Omega}^{i}(l^{i}, b^{i}, d^{i}) = (\Omega_{l}^{i}, \Omega_{b}^{i}, \Omega_{d}^{i})$ be the spin vector of the $i$-th galaxy with galactic longitude $l^{i}$, \rm galactic latitude $b^{i}$, \rm and distance $d^{i}$. \rm  The distance $d^{i}$ to the galaxy can be approximated for the nearby universe by $d^{i} = c z^{i} /H_{0}$, \noindent where $c$ is the speed of light, $z^{i}$ is the spectroscopic/photometric redshift of the $i$-th galaxy and $H_{0}$ is the Hubble constant.

Measuring the 3D vector $\mathbf{\Omega}^{i}$ is nontrivial, but one  can easily judge the sign $h^{i}$ of $\Omega_{d}^{i}$. This can be done by examining the spiral winding sense to see if it is S-wise ($h^{i}=-1$) or Z-wise ($h^{i}=+1$), since all spiral galaxies can be assumed to be trailing \citep[][: Paper I]{Iye2019}.  The net effect of misidentifying S-spirals vs Z-spirals and improbable presence of leading spiral arms would be reduction of the dipole strength, should it exist. For simplicity, hereafter we assume that all the $\mathbf{\Omega}^{i}$ have an unit length, namely $\Omega_{d}^{i}=h^{i}$.

By compiling the spin parity, $h^{i}=\pm 1$, together with the coordinates of spiral galaxies from image archives such as SDSS, Pan-Starrs1, ESO-DR2, DES, Subaru HSC, and others, we estimate that one can produce a spin parity catalog of up to a million spiral galaxies in the 3D volume within 1 Gpc of the Earth.   We are developing an analysis scheme to probe any partial volume within this volume, not necessarily centered on the Earth. This will be discussed in a forthcoming paper. For the moment, however, we consider a local volume centered on the Earth.

To analyze the spin vector distribution in terms of spherical harmonics expansion, one can first examine on its dipole component $Y_{1}^{0}$, quadrupole component $Y_{2}^{0}$ and higher components.

If we define a unit vector to a fiducial pole $\mathbf{P}(l_{P}, b_{P})$, one can calculate the inner product
\begin{equation}
D(l_{P}, b_{P})=\sum_{i=1}^{N}{h^{i}\mathbf{\Omega}^{i}}\mathbf{P}/N=\sum_{i=1}^{N} {h^{i}\cos\theta^{i}}/N
\end{equation}
\noindent where the angle $\theta^{i}$ is the angle between the direction of the $i$-th galaxy and the direction of the fiducial pole vector $\mathbf{P}$. By determining the direction of the vector $\mathbf{P}$ for which the amplitude $D_{max}=|\mathbf{D}(l_{P}, b_{P})|$ takes the largest value, one can derive the dipole vector $\mathbf{D}_{max}$ of the observed distribution.  

In fact, $D_{max}$ can be obtained simply by calculating a vector
$\mathbf{G}$, which is the vector sum of the unit radial vectors pointing to
the direction of the $i$-th galaxy multiplied by the helicity $h^{i}$.
The inner product of $\mathbf{G}$ with $\mathbf{P}$ takes the maximum value when $\mathbf{P}$ is
pointing in parallel to $\mathbf{G}$, and then $D_{max}$ is $\|\mathbf{G}/N\|$.

\subsection{Perfect Dipole Distribution}

We assume an extreme case of perfect dipole segregation, where all S-spirals are in the northern hemisphere and all Z-spirals are in the southern hemisphere as shown in Figure~\ref{fig2}(a). The weight factor $\cos\theta^{i}$ of Equation 2 shows that the galaxies toward the dipole axis with small $\theta^{i}$ add to $D_{max}$ while those near the equator with $\theta^{i}\sim\pi/2$  reduce $D_{max}$.  It is easy to see by integration that the expected mean amplitude $D_{max}^{perfect}$ of all sky sampling is 0.5. Perfect but random dipole distribution produces fluctuation around this expected mean amplitude of 0.5. Monte Carlo simulation shows that the associated standard deviation from this mean amplitude for perfect random dipole distributions is $\sim0.3/\sqrt{N}$.

\subsection{Effect of Non-uniform Sky Coverage for a Perfect Dipole Distribution}

Although the currently available data from imaging surveys are growing rapidly, the data do not yet fill the entire sky uniformly. There may be observational biases that affect evaluation of S/Z number asymmetry and/or the evaluation of the observed dipole vector in the S/Z distribution. Let us examine the effect of non-uniform sky coverage for the perfect dipole distribution case mentioned in the previous subsection.

When the sky sampling is limited to a certain narrow cone direction, the dipole amplitude can take any value in the range $0\le D_{max}\le1$ depending on the direction to the sample. The largest amplitude $D_{max}=1$ is obtained when the sampling is toward the pole, $\theta=0$ or $\pi$, and the lowest amplitude $D_{max}=0$ is observed toward the equator $\theta=\pi/2$. Therefore depending on the direction of the biased small sky sampling, the resulting $D_{max}$ can be larger or smaller than the whole sky sampling.

\begin{figure}[ht]
\centering
\includegraphics[scale=0.85]{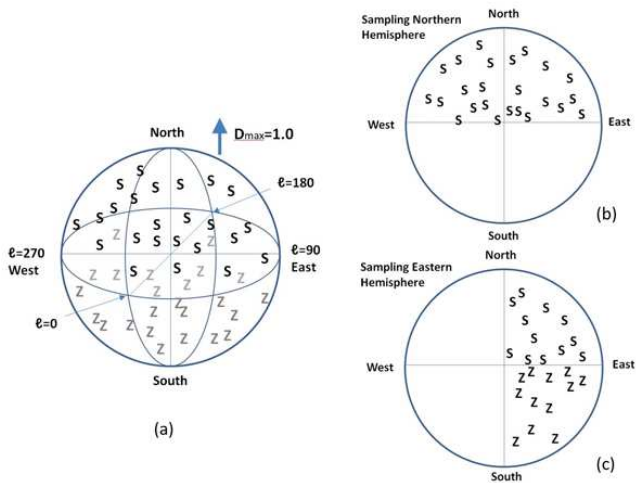}
\caption{Effect of non-uniform sky coverage on the number asymmetry and on dipole evaluation. (a) Distribution of spiral galaxies for an extreme case of $100\%$ dipole asymmetry. All spirals in the northern hemisphere are exclusively S-spirals, while all spirals in the southern hemisphere are exclusively Z-spirals. (b) If the sampling volume is limited only to the northern hemisphere, then only S-spirals are observed, showing an extreme asymmetry in the number count.  The dipole amplitude, however, is equal to that derived for the entire sky. (c) If the sampling volume is limited to the eastern hemisphere, the dipole amplitude is equal to that of the entire sky and no asymmetry is observed in the number count.}
\label{fig2}
\end{figure}

Consider a non-uniform sampling of the complete dipole distribution covering only the northern hemisphere as shown in Figure~\ref{fig2}(b). There would be a conspicuous number count asymmetry. However, the dipole strength observed would be equal to that of the whole sky sampling. If the sampling were limited to the eastern hemisphere as shown in Figure~\ref{fig2}(c), there would be no number asymmetry.  The dipole amplitude, again, would be equal to that of full sky sampling.

The number asymmetry and the dipole amplitude thus provide complementary information to analyze large scale symmetry-breaking in the spin distribution of galaxy ensembles.

\subsection{Random Flight Simulation for an Isotropic Distribution}

For a uniformly randomly distributed set of $N$ vectors with $\mathbf{\Omega^{i}}$, the resultant vector sum $\mathbf{D_{max}}$ will have an isotropic distribution with non-zero amplitude, unless the summation has incidentally completely  canceled $\mathbf{D_{max}}$. 

As $D_{max}$ can be calculated by $\|\mathbf{G}/N\|$, our problem is equivalent to a well-studied mathematical problem, 3D random walk (random flight). Namely, it is equivalent to find the distribution of distance to the final point vector $\bf{R}$ of a particle that starts from the origin and takes $N$ steps of a 3D random walk.
One can evaluate the mean amplitude of $D_{max}$
and the standard deviation around the mean amplitude
as a function of $N$.

\citet{Chandrasekhar1943} established that the probability density function $W(\bf{R})$ of the final displacement vector $\bf{R}$ of random flights for a large ${N}$ will be a 3D Gaussian distribution.    
\begin{equation}
W({\bf{R}})=\frac{1}{(2\pi N\langle r^{2}\rangle_{Av}/3)^\frac{3}{2}} exp(-3|{\bf{R}}|^{2}/2N\langle r^{2}\rangle_{Av})
\end{equation}
\noindent where $\langle r^{2}\rangle_{Av}$ is the expected mean square displacement, which is in our case $\langle r^{2}\rangle_{Av}=1$.

The distribution of $D_{max}^2$, therefore, follows the chi-squared distribution for three degrees of freedom. The distribution of $D_{max}$, hence, follows the chi distribution\footnote{Weisstein, Eric W. "Chi Distribution." From MathWorld--A Wolfram Web Resource. https://mathworld.wolfram.com/ChiDistribution.html}, a square root of chi- squared distribution. The expected mean amplitude $\overline{D}_{max}$ is
\begin{equation}
\overline{D}_{max} = \frac{\sqrt{2}\Gamma(2)}{\sqrt{3N}\Gamma(3/2)}=\frac{2\sqrt{2}}{\sqrt{3\pi N}} \sim \frac{0.921}{\sqrt{N}}. 
\end{equation}

The associated standard deviation from this expected mean distance is given by 
\begin{eqnarray}
Stddev&=&
\sqrt{\frac{2[\Gamma(3/2)\Gamma(5/2)-\Gamma(2)^{2}]}{3\Gamma(3/2)^{2}N}}
=\sqrt{\frac{3\pi-8}{3\pi N}}\nonumber\\
&\sim& \frac{0.389}{\sqrt{N}}.
\end{eqnarray}

We can use these formulae to evaluate the statistical significance $\sigma_{D}$ of the observed spin dipole strength $D_{max}$ in any ensemble of spiral galaxies using
\begin{equation}
\sigma_{D} \sim (D_{max}\sqrt{N}-0.921)/0.389. \rm
\end{equation}

\subsection{Detectability of Dipole Component}

Consider an ensemble of galaxies where a perfect dipole system and  a uniform random system are mixed with fractions $p$ and $1-p$, respectively.
The observed dipole vector $\mathbf{D_{max}}$ would be, on average, a vector sum of $\mathbf{D_{max}^{perfect}}$ with an amplitude $0.5p$ and a random vector with an amplitude $(1-p)\times0.921/\sqrt{N}$ in an arbitrary direction.
To discern the intrinsic dipole from the random dipole with a statistical significance of $s$-sigma, the following relation is required
\begin{equation}
0.5p \ge (s+1)*0.921(1-p)/\sqrt{N}
\end{equation}

This implies
\begin{equation}
N\ge (\frac{0.921(1-p)(1+s)}{0.5p})^{2}
\end{equation}

The Equation 8 indicates that between one hundred thousand or one million spirals are necessary to detect a 3\%, or 1\% residual inherent dipole system at $5\sigma$ confidence level, respectively.

Real data are not always obtained uniformly. The quantitative discussion of the non-uniform sampling of a random distribution in the general case is not straightforward. We examine, however, the actual S/Z data of SDSS spirals in the next section and compare the observed dipole amplitude with those expected from simulated random distributions.

\section{Application to the S/Z Distribution of SDSS Spiral Galaxies}

\subsection{Reanalysis of Shamir's Spin Catalog}
To apply our formulation of dipole anisotropy to real data, we studied two samples using
Shamir's catalog based on SDSS photometric data \citep{Shamir2017a}\footnote{https://data-portal.hpc.swin.edu.au/dataset/data-for-galaxy-assymetry-experiment}. 
The first sample retains all 162,516 spirals from \bf{the} original catalog. The original sample shows an S/Z dipole signal of $D_{max}=0.00489$, with its axis pointing toward $(l, b)=(189^{\circ}, +15^{\circ})$.  This axis coincides with that reported in \citet{Shamir2020a}, ($\alpha = 88^{\circ}, \delta = +36^{\circ}$), which is ($l = 175^{\circ}, b = +5^{\circ}$) for an SDSS sample, considering the $1\sigma$ estimation error of about $30^{\circ}$ in both coordinates. 

For calibration, we made 50,000 independent Monte Carlo simulations by assigning $h^{i}=\pm1$ randomly to the 162,516 spirals and measured the simulation's $\mathbf{D_{max}}$. The resulting $\mathbf{D_{max}}$ shows an isotropic distribution with an ensemble mean amplitude of $D_{max}=0.00225$ and an associated amplitude standard deviation of $\sigma=0.00105$, as shown in Table 1. The observed $D_{max}$ from the original catalog, therefore, has an amplitude, that is larger than the mean amplitude by $\sigma_{D}=2.52$. \rm
The number dominance of Z-spirals over S-spirals here is $\sigma_{N}=4.89$.

The second sample we studied is a volume-limited sample retaining 111,867 spirals with measured redshift in the range $0.01\le z \le 0.1$.  To avoid any possible effect of local peculiar motions, 162 nearby spirals at $z \le 0.01$ were removed from this volume-limited sample.  This sample shows a stronger S/Z dipole signal $D_{max}=0.00773$ with its axis pointing toward $(l, b)=(138^{\circ}, -38^{\circ})$.  

\begin{figure}[ht]
\centering
\includegraphics[scale=0.55]{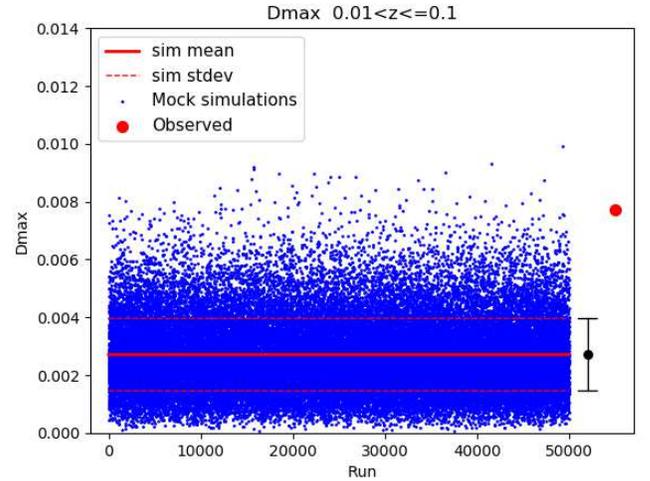}
\caption{Dipole amplitude $D_{max}$ measured in 50,000 mock simulations and in a sample of 111,867 spirals within a volume of $0.01 \le z \le 0.1$ from \citet{Shamir2017a}. The solid line shows the mean dipole amplitude expected from the simulation and the broken line shows $\pm 1 \sigma$ deviation from the mean amplitude. The observed dipole (red circle) for this volume-limited sample is $4.00\sigma$ away from the expected mean amplitude.}
\label{figure_3}
\end{figure}

The ${D_{max}}$ value observed for this sample, together with that of 50,000 simulated mock samples is shown in Figure~\ref{figure_3}.  The observed $D_{max}$ from the original catalog limited to a volume defined by redshift, therefore, has an amplitude, that is larger than the mean amplitude by $\sigma_{D}=4.00$.\rm This amplitude happens to be close to the value $4.34\sigma$ reported in \citet{Shamir2020a}. The number dominance for this sample is $6.36\sigma$.
\vspace{2mm}

\vspace{2mm}
\begin{figure}[ht]
\centering
\includegraphics[scale=0.55]{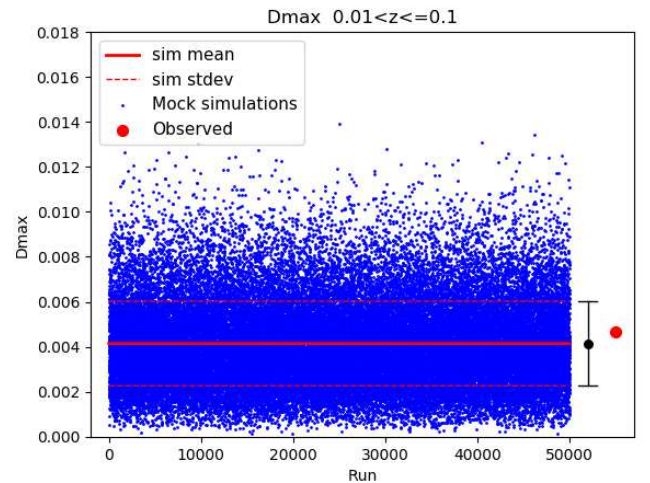}
\caption{Same as Figure 3 but for a 'cleaned' sample of 48,089 spirals, removing multiple entries in the original catalog.}
\label{figure_4}
\end{figure}

\subsection{Analysis of New Cleaned Catalog}

Upon re-examining Shamir's original catalog, we found significant duplication of entries in its tables. Apparently, \citet{Shamir2017a} used PhotoObjAll to search the SDSS catalog. According to the Table Description of DR8\footnote{http://skyserver.sdss.org/dr8/en/help/docs/tabledesc.asp}, \rm the view of PhotoPrimary, instead of PhotoObjAll, should be used to avoid duplicate detections.  PhotoObjAll returns every entry from the searched images without checking for duplication of the same object.  

In fact, 34,198 spiral galaxies were found to have multiple entries with their coordinates within 3 arcsec distance of each another. For instance, one of the galaxies in his catalog, SDSS J031945.63-000437.9 (objid=1237666300555427954) at $z=0.037$, has 89 separate entries. We confirmed by visual inspection that all of them are the same galaxy. 

A total of 106 spirals had contradicting S/Z classification among the duplicated entries. We rectified these contradictions by assigning a spin parity via visual inspection in some cases. In other cases, where the voting was significantly split, we adhered to the outcome of majority voting.

After removing the duplicated entries, the number of spiral galaxies is reduced to 72,888, that is $45\%$ of the original number.  The number of spirals in the volume-limited sample  range $0.01 \le z \le 0.1$ is 48,089 (23,819 S-spirals and 24,270 Z-spirals), 43\% of the 111,867 counted in the original catalog. The measured amplitudes for these 'cleaned' samples with and without the volume limitation are summarized in Table 1.

The observed dipole for the entire cleaned sample set has $D_{max}=0.00535$ with its axis pointing toward $(l, b)=(192^{\circ}, +79^{\circ})$.  50,000 random mock simulations for the cleaned sample shows a mean amplitude $D_{max}=0.00336$ and a standard deviation $\sigma=0.00154$. Therefore, the observed dipole deviates from the expected mean strength only at a level of $\sigma_{D}=1.29$. Number dominance of Z-spirals over S-spirals is also only at the $\sigma_{N}=1.87$ level.

The observed dipole for the cleaned sample limited to the redshift range $0.01 \le z \le 0.1$, is $D_{max}=0.00468$ with its axis pointing toward $(l, b)=(106^{\circ}, +57^{\circ})$.  Results of 50,000 random mock simulations for the cleaned sample shows a mean amplitude $D_{max}=0.00414$ and a standard deviation $\sigma=0.00188$, as shown in Figure~\ref{figure_4}. 

Therefore, the observed dipole deviates from the expected mean strength only at the $\sigma_{D}=0.29$ level. Column (13) of Table 1, obtained from Equation 6 for uniform sampling is not necessarily correct for non-uniform sampling. However, the fact that Column (13) shows values close to the values in column (12), obtained from actual sampling, suggests their relevance. Finally, the number dominance of Z-spirals over S-spirals is also only at the $\sigma_{N}=2.06$ level.

Comparison of Figures~\ref{figure_3} and ~\ref{figure_4} clearly shows that the apparent pseudo dipole signal observed in Figure~\ref{figure_3} came from massively duplicated data in the original catalog.

As a final remark, \citet{Longo2011} made a similar analysis of his sample of 15,158 spirals with $z\le 0.085$ from SDSS DR6. He used the terms left-handed (S-spiral in the present paper) and right-handed (Z-spiral).  He found a dipole asymmetry by plotting the distribution of $A=(Z-S)/(Z+S)$. The dipole strength under his definition was $-0.0408\pm0.011$, found with an axis pointing at $(l, b)=(52^{\circ}, +68.5^{\circ})$ \rm at $5.6\sigma$. The relation of his study to the current work has yet to be investigated.

\begin{table*}[]
\begin{center}
\caption{Observed dipole asymmetry in the spin distribution of SDSS galaxies. Column(6) is the number asymmetry significance level. Columns (7-9) shows the observed amplitude and the direction of the dipole $\mathbf{D_{max}^{obs}}$, while Columns (10) and (11) show the mean amplitude of $D_{max}$ from 50,000 mock simulations and the standard deviation from the mean value, respectively.  Column (12) is the observed significance level of $D_{max}$.  Column(13) is the expected significance level for an isotropic random distribution as given by Equation 6.}
\begin{tabular}{|c|c|c|c|c|c|c|c|c|c|c|c|c|} \hline
(1)&(2)&(3)&(4)&(5)&(6)&(7)&(8)&(9)&(10)&(11)&(12)&(13)\\
Sample&redshift range&N&N(S)&N(Z)&$\sigma_{N}$&$D_{max}^{obs}$&$l$&$b$&$\overline{D}_{max}$&stddev&$\sigma_{D}^{obs}$&$\sigma_{D}^{iso}$ \\ \hline
Original&All&162,516&80,272&82,244&4.89&0.00489&189&+15&0.00225&0.00105&2.52&2.70\\
Original&$0.01\le z \le 0.1$&111,867&54,870&56,997&6.36&0.00773&138&-38&0.00271&0.00126&4.00&4.28\\
Cleaned&All&72,888&36,191&36,697&1.87&0.00535&192&+79&0.00336&0.00154&1.29&1.34\\
Cleaned&$0.01 \le z \le 0.1$&48,089&23,819&24,270&2.06&0.00468&106&+57&0.00414&0.00188&0.29&0.27\\
\hline
\end{tabular}
\end{center}
\end{table*}

\section{Conclusion} \label{sec:discussion}

We present a formulation to quantify the observed dipole amplitude $D_{max}$ of the S/Z spin parity distribution of spirals in an ensemble of galaxies.  We show that the statistical significance $\sigma_{D}$ of this quantity can be calibrated with that expected from 3D random flight simulations. 

Notably, we found that the S/Z spin catalog published by \citet{Shamir2017a} contains a significant amount of duplicated data, which at least partly caused the increased level of dipole asymmetry that we observed.  After removing the duplicated entries from the catalog, we found that the distribution is compatible with random distribution.   We conclude that the SDSS sample of spiral galaxies does not show large scale anisotropy in the spin distribution of galaxies.

Recently, \citet{Shamir2020b} studied another dataset based on spectroscopic sample to detect a significant dipole signal.  Previously \citet{Shamir2012} also presented that a dataset based on SDSS SpecObj shows a significant dipole signal, while the direction of the axis is different.  The different results from the different datasets will be investigated elsewhere.

The current authors are compiling a large coherent dataset of spiral winding evaluation as derived from several modern large image datasets, including the  SDSS \citep{Aguado2018}, the Panoramic Survey Telescope and Rapid Response System \citep[Pan-Starrs1:][]{Chambers2016}, the Hyper Suprime-Cam \citep[HSC:][]{Miyazaki2018} of Subaru Telescope, and the Dark Energy Survey \citep{Abbott2018} by deep learning algorithm. \citet{Tadaki2020} developed a deep learning algorithm to judge S/Z winding of 76,635 spiral galaxies from the Hyper Suprime-Cam Subaru Strategic Program Data Release 2 dataset \citep[HSC SSP DR2:][]{Aihara2019} sample, with further objective to generate a galactic spin data catalog.


\acknowledgments

Special thanks are due to Lior Shamir who gave us useful comments to improve the present paper. 
We thank the anonymous referee for helpful comments.
Data analysis of this study was in part carried out on the
Multi-wavelength Data Analysis System operated by the Astronomy Data
Center (ADC), National Astronomical Observatory of Japan.
This research has made use of NASA's
Astrophysics Data System Bibliographic Services.
Funding for SDSS-III has been provided by the Alfred P. Sloan
Foundation, the Participating Institutions, the National Science
Foundation, and the U.S. Department of Energy Office of Science. The
SDSS-III web site is http://www.sdss3.org/.
SDSS-III is managed by the Astrophysical Research Consortium for the
Participating Institutions of the SDSS-III Collaboration including the
University of Arizona, the Brazilian Participation Group, Brookhaven
National Laboratory, Carnegie Mellon University, University of Florida,
the French Participation Group, the German Participation Group, Harvard
University, the Instituto de Astrofisica de Canarias, the Michigan
State/Notre Dame/JINA Participation Group, Johns Hopkins University,
Lawrence Berkeley National Laboratory, Max Planck Institute for
Astrophysics, Max Planck Institute for Extraterrestrial Physics, New
Mexico State University, New York University, Ohio State University,
Pennsylvania State University, University of Portsmouth, Princeton
University, the Spanish Participation Group, University of Tokyo,
University of Utah, Vanderbilt University, University of Virginia,
University of Washington, and Yale University.

\end{document}